\documentclass[sigconf]{acmart}
\renewcommand\footnotetextcopyrightpermission[1]{} % removes footnote with conference information in first column

\usepackage{graphicx,float,subfigure,amsmath,algorithm}
\usepackage[noend]{algpseudocode}
\usepackage{multirow}
\usepackage{float}
\AtBeginDocument{%
 \providecommand\BibTeX{{%
 \normalfont B\kern-0.5em{\scshape i\kern-0.25em b}\kern-0.8em\TeX}}}

\begin{document}

\title{Predicting the Biological Classification of Cell-Cycle Regulated Genes of {\it Saccharomyces cerevisiae}  using Community Detection Algorithms on Gene Co-expression Networks}

\author{ Jhoirene Clemente,  Gabriel Besas,  Jerick Callado,  \and John Erol Evangelista}
\email{cgbesas@up.edu.ph, jccallado@up.edu.ph, jbclemente@up.edu.ph, jmevangelista@up.edu.ph}
\affiliation{%
 \institution{Algorithms and Complexity Lab, 
Department of Computer Science, \\
University of the Philippines Diliman}
 \city{Quezon City}
 \country{Philippines}
 \postcode{1101}
}


\begin{abstract}
The conventional approach for analyzing gene expression data involves clustering algorithms. Cluster analyses provide partitioning of the set of genes that can predict biological classification based on its similarity in $n$-dimensional space. 
In this study, we investigate whether network analysis will provide an advantage over the traditional approach. We identify the advantages and disadvantages of using the value-based and the rank-based construction in creating a graph representation of the original gene-expression data in a time-series format. We tested four community detection algorithms, namely,  the Clauset-Newman-Moore (greedy), Louvain, Leiden, and Girvan-Newman algorithms in predicting the 5 functional groups of genes.  We used the Adjusted Rand Index to assess the quality of the predicted communities with respect to the biological classifications. We showed that Girvan-Newman outperforms the 3 modularity-based algorithms in both value-based and ranked-based constructed graphs. Moreover, we also show that when compared to the conventional clustering algorithms such as $K$-means, Spectral, Birch, and Agglomerative algorithms, we obtained a higher ARI with Girvan-Newman.  This study also provides a tool for graph construction, visualization,  and community detection for further analysis of gene expression data.

\end{abstract}




\maketitle
\pagestyle{plain}
\section{Introduction}
Graph analytics has brought significant advances to our understanding of complex systems. One interesting graph substructures are communities or clusters which are identified as a tightly knit set of vertices.  Community detection is used in various fields including, biological, social, technological, and information networks \cite{Fortunato2010}.  In Biological networks, these communities could be a group of genes or proteins that interact highly with each other.  This study will focus on finding communities in gene co-expression networks.  Specifically,  this research will investigate how the graph will be built from yeast gene expression data and how feasible it is to predict 5 the functional group of genes by using a graph-based approach instead of the conventional clustering algorithms. 

We use time-series gene expression data from yeasts created by Cho et al. \cite{Cho1998}. This data is modified by Yeung \cite{Yeung2001} to create a part of the benchmark dataset for validating clustering algorithms for gene expression data. In our previous work, we used the same data set for analyzing two visualization techniques in \cite{Clemente2011} and we found that functional groups of genes and their relationships reflect in the Nonmetric-multidimensional scaling (NMDS) visualization. We would like to explore further if the functional groupings of genes are reflected by using another powerful combinatorial model,  which is graphs. 

We created several graph representations of the yeast gene expression data.  We explored two graph construction algorithms, namely the rank-based and value-based construction.  Most co-expression networks are obtained using value-based graph construction \cite{Ruan2010}. To name a few, this method is used in creating a gene co-expression network for the human genome \cite{Jordan2004, Lee2004, Elo2007}.  Value-based construction is also used to create a gene expression network used to compare human and chimpanzee brains \cite{Oldham2006}. It is also used to study several genes related to chronic fatigue syndrome (CFS) in \cite{Presson2008}, and mouse gene related to weight \cite{Ghazalpour2006}.  A study by Ruan et al. \cite{Ruan2010} argues that the rank-based approach claims to better capture the global topology of biological systems which involved both strongly and weakly co-expressed modules unlike the modules obtained by value-based construction.  A couple of gene co-expression networks that are obtained using rank-based approaches includes gene co-expression networks that link cardio-vascular disease and Alzheimer's \cite{Ray2008}. Rank-based graph construction is also used to create a network to analyze cellular pathways that regulate different biological processes involved in gastric cancer.  Another example is used to analyze genes from humans, flies, worms, and yeast to discover 163 genes that are conserved across evolution \cite{Stuart2003}.

We obtained several graph representations using the value-based and the rank-based constructions and subjected each to a couple of community detection algorithms.   Community detection is an NP-hard problem. Moreover, complex networks naturally have a high number of vertices, so using exact approaches for community detection is impractical. In this study, we used four heuristic algorithms.  The greedy,  Louvain, and Leiden algorithm uses modularity to measure the quality of the community structure.  On the other hand,  the Girvan-Newman approach uses a heuristic that iteratively removes valuable edges to reveal the communities in the graph.  We used the biological classification of the genes to create a baseline truth about the intended communities. We compare the predicted communities of the algorithms using the  Adjusted Rand Index (ARI).

\section{Gene Expression Data}
Genes are segments in the DNA that code for a specific biological function. A particular gene is expressed whenever a copy of it is transcribed and later on translated into proteins to carry out a specific biological process. Microarray technology allows measuring thousands of gene expression levels at once.

In this study, we used the gene expression data of \emph{Saccharomyces cerevisae} or commonly known as yeast from \cite{Spellman1998}. The original data consists of 6,220 gene expression levels taken into 17-time points. The duration of the interval covers two cell cycles of a synchronized sample of yeasts. From the set of monitored genes, 416 of these genes are found to exhibit cell cycle dependent periodicity \cite{Cho1998}. The work of Cho et al. \cite{Cho1998} also provided a characterization of genes in terms of their involvement in cell cycle regulation. Yeung et al. \cite{Yeung2001} provided a subset of genes, where no two genes belong to more than one phase. A subset of 384 genes is grouped according to the 5 phases of the cell cycle. Each group corresponds to a specific cell cycle phase which activates in the following order: early G1 phase, late G1 phase, synthesis phase, second growth phase, and mitosis or cell division. In this paper, we refer to the groupings as Group 1 to 5, respectively. We will use the term functional group to refer to the group of genes belonging to a common function. Unless specified otherwise, the main functional grouping of genes is based on the 5 phases of the cell cycle. A more specific grouping was provided in the work of Cho et al. \cite{Cho1998}, where they provided biological characterization to some of the genes belonging to a single functional group. Examples of such characterizations are genes involved in cell cycle regulation, directional growth, DNA replication, mating pathway, glycolysis/respiration, biosynthesis, and transcription factors. 

\begin{figure}[h]
\centering
\includegraphics[width=0.48\textwidth]{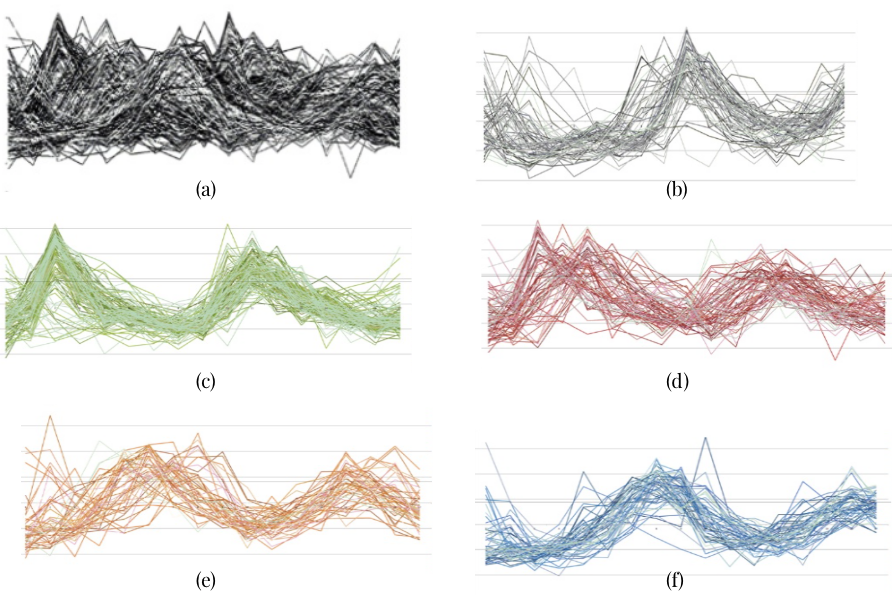}
\caption{(a) Expression levels of 384 genes accross the 17 time points. The genes are classified into five groups corresponding to the five phases of the cell cycle, i.e., (b) Early G1 phase (c) Late G1 phase (d) Synthesis phase (e) Second growth phase and (f) Mitosis. }
\label{fig:expression_level}
\end{figure}

We used a normalized version of the gene expression data from \cite{Tamayo1999}. The expression levels are normalized to have a mean equal to 0 and a variance equal to 1. In Figure \ref{fig:expression_level}, we show the expression level of genes belonging to the five different functional groups. 
For each group, we see how the expression level of genes fluctuates through time. We refer to the total number of genes as $n = 384$ and the total number of time points as $t=17$. The data is initially represented as an $(n \times t)$ matrix, where each row represents a gene vector of size $t$. 

\begin{figure}[h]
\centering
\includegraphics[width=0.48\textwidth]{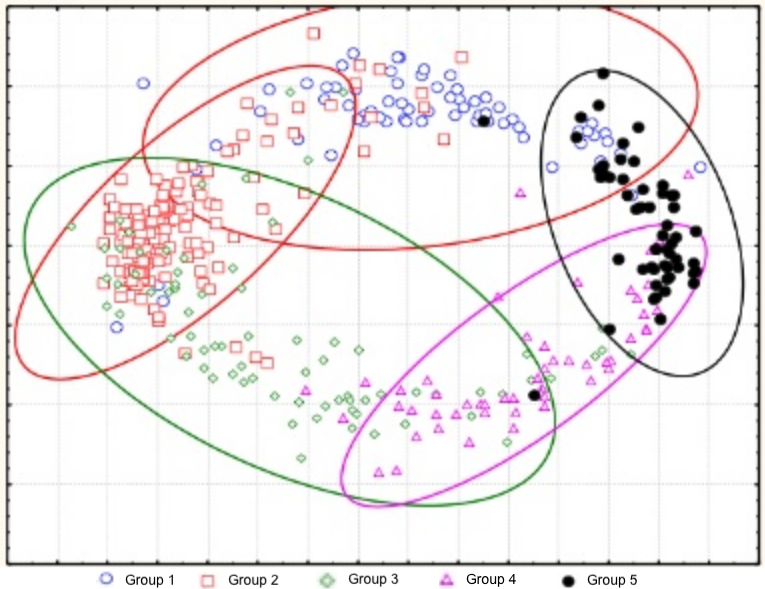}
\caption{Non-metric multidimensional scaling of the yeast gene expression data in \cite{Clemente2011}. Each point in the scatterplot is a gene and the colors represent the biological classification. Fitted confidence ellipses are displayed to show how biological function prediction can be made by visual closeness to the five different groups. }
\label{fig:nmds}
\end{figure}

In our previous study, we investigate the use of vector fusion and non-metric multidimensional scaling (NMDS) visualization of the 384 genes \cite{Clemente2011}. We used the Euclidean distance as a metric for gene similarity. In Figure \ref{fig:nmds}, we show the NMDS visualization of the gene expression data in a 2-dimensional view. The resulting non-metric visualization shows visual clustering of the five classifications of genes. Moreover, the 2D representation captures the temporal characteristics of the phases of the cell cycle. In counterclockwise, we can see the visual proximity of the groups starting from Group1 to Group 5, which activates in order according to the cell cycle phase. Further investigation of the use of the 2D visualization in predicting functional group of genes is studied in the works of \cite{Salido2012} and \cite{Salido2016a}. 

This work is an extension of the previous studies in representing gene relationships and predicting functional groups of genes. We extended the 2D visualization in \cite{Clemente2011} to graph representations to capture additional information about the data which can be used in predicting the biological function of genes. In this following section, we will describe how we model the yeast gene expression data using a gene co-expression network. 

%		n this research, we are trying to create a graph for the co-expression data. We are planning to achieve this by measuring the correlation of each gene to another and creating edges between them depending on the correlation. The problem we are addressing in the pre-processing is the expression data of the genes does not have a specific range. The amount of percentage increase and decrease for each gene will vary and comparing each gene expression data in its raw state may not encapsulate its relations with each other. By normalizing the data, we aim to compare the similarity of periodicity of gene expressions and produce a more suitable visualization for the graph.

\section{Graph Representation and Visualization}

In our graph representation, each vertex is a gene and there is an edge connecting two genes if they are co-expressed. We formalize the concept of gene co-expression by utilizing Pearson's correlation as a similarity metric. Since this metric is symmetric, the resulting graph is an undirected graph with edge weights equal to the correlation value. In this paper, we use the terms network and graph interchangeably.

In the following subsections, we will detail the resulting graphs obtained by varying these input parameters. We used two different graph construction algorithms, namely, \emph{value-based} and \emph{rank-based} construction. The main difference between the two algorithms is how the edges are formed in the graph. 

To visualize the resulting graph representations, we used the Fructherman-Reingold algorithm \cite{Fruchterman1991}. The algorithm computes for the positions of the nodes in 2D or 3D space by simulating a physical system or rings and springs. Each node is represented by a ring, while the edges represent the springs. The weights associated with the edges represent the attractive forces between the rings. The final position of the rings in space generally reflects the similarity between different nodes with respect to the edge-weights. 

A force-directed layout such as the Fruchterman-Reingold algorithm reveals the underlying topology of a network and is said to be suitable for highly modular networks with distinct communities \cite {Koutrouli2020}. The Fruchterman-Reingold algorithm is a widely used force-directed algorithm for several gene co-expression networks in the literature. For instance, this algorithm is proven to be relevant in extracting biologically significant modules of pig genes \cite{Villa-Vialaneix2016}. This technique is also used to identify a subnetwork of yeast genes that are activated in response to its adaptation to different environments \cite{Campiteli2009}.

\subsection{Value-based Graph Construction}

%Formally, the graph representation of the gene expression data is an undirected and unweighted graph $G(V,E)$, where $V$ is the of all genes with$ |V|=n$, $E$ is the set of all edges connecting $u$ and $v$ if $\rho(u,v)>= \delta$, for some $\delta$. 
We create a corresponding graph representation of the yeast gene-expression data using value-based construction. To create a gene co-expression network, we compute the pairwise similarity of genes using Pearson's correlation coefficient. An input parameter $\delta$ is used to identify whether two genes are co-expressed. 

Formally, let $G=(V, E)$ be the corresponding co-expression network obtained from the yeast gene-expression data. The resulting graph $G$ is a vertex labeled graph with $|V| = n = 384$. In the value-based construction, there is an edge connecting two vertices $u$ and $v$ if and only if the Pearson's correlation coefficient $\rho(u, v) \geq \delta$, for some correlation threshold $\delta$. 

We created several graph representations by varying the input parameter $\delta$. In total, we obtained 5 graph representations of the same data set by creating a corresponding graph for $\delta = \{0.70, 0.75, \ldots, 0.95\}$. We summarized the properties of the graphs in Table \ref{table:vbnet}.

\begin{table}[h]
\resizebox{0.9\width}{!}{
\begin{tabular}{ | c | c c c c c c |}
\hline
$\delta$	&	0.70	&0.75	&0.80	&0.85&	0.90 &	0.95 \\ \hline
Edges	&	9748	&	7383 &	5058&	2925	&1177	&154\\
Singletons	&	3	&12&	26&	68&	142	&299 \\
Number of CC&	3 &	2	&3	&3&	18	&11 \\
Largest CC	& 375	&369&	354	&310&	116&	53 \\
%Clustering coefficient & 0.706	& 0.647	& 0.593	& 0.446	& 0.244	&0.053 \\
\hline
\end{tabular}}
\caption{Summary of network properties for graphs obtained using value-based constrution}
\label{table:vbnet}
\end{table} 

\begin{figure}[h]
\centering
\includegraphics[width=0.48\textwidth]{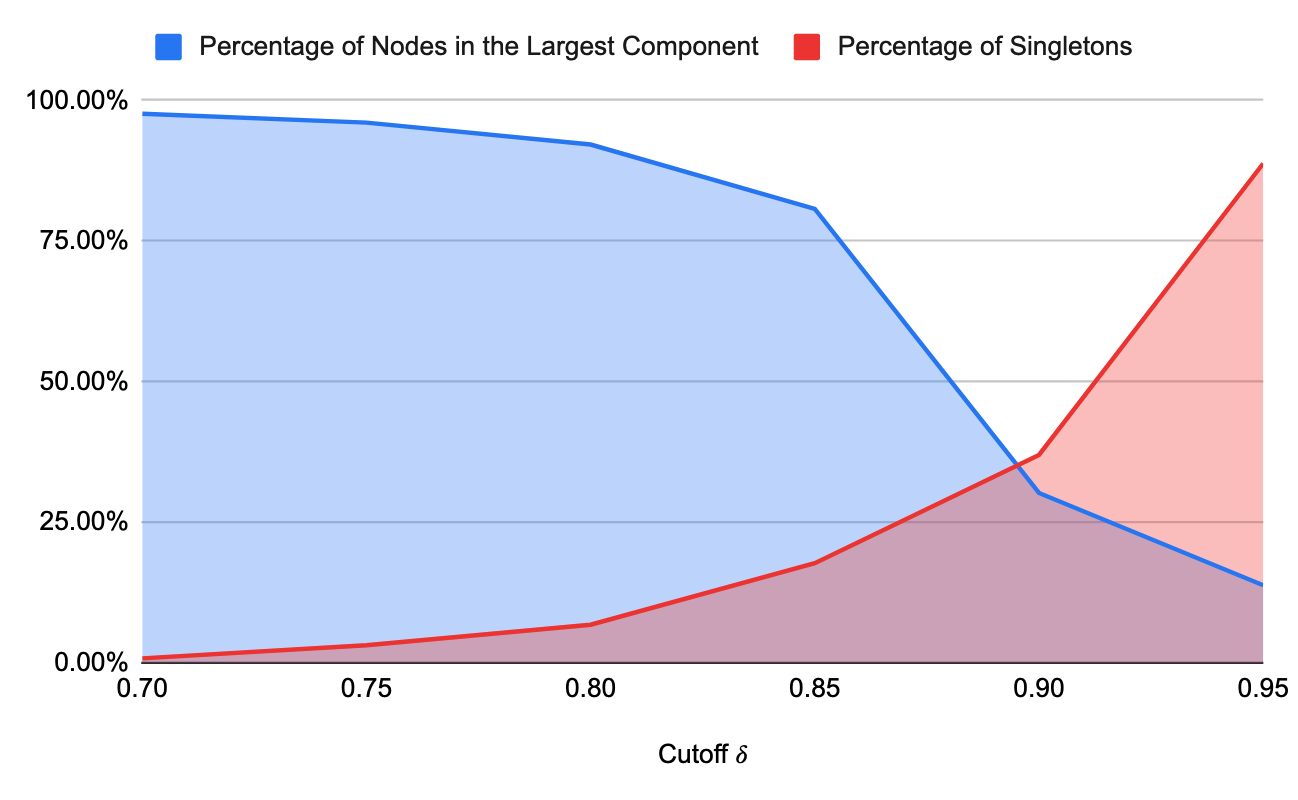}
\caption{Percentages of nodes that are part of the largest component in $G$ and the percentage of nodes that are singletons as the threshold $\delta$ increases. }
\label{fig:vbcc}
\end{figure}

In Table \ref{table:vbnet}, we can see how the total number of edges decrease as we increase the parameter $\delta$. In Figure \ref{fig:vbcc}, we show the percentage of nodes that are part of the largest connected component and the corresponding percentage of singletons in the resulting graph as we increase the parameter $\delta$. The percentage of the number of nodes that is part of the largest connected component is inversely proportional to the total number of singletons. 

The graph obtained with $\delta=0.95$ have edges connecting highly correlated yeast genes. This high threshold results in creating a graph with the lowest edge count and the highest number of singletons. In this graph, about $78\%$ are singletons. In the context of community detection, we cannot get information about interesting relationships amongst these genes. On the other hand, this graph representation revealed 11 non-trivial connected components with the largest connected component having 53 nodes. 9 out of 11 connected components consist of genes belonging to the same functional group. One connected component consists of a transcription factor `YDL197c' in the 3rd functional group and a gene involved in repair and recombination `YLR383w' in the 2nd functional group. The largest connected component has 49 genes belonging to the 2nd functional group while 4 genes are involved in the 3rd functional group. 

\begin{figure}[h]
\centering
\includegraphics[width=0.48\textwidth]{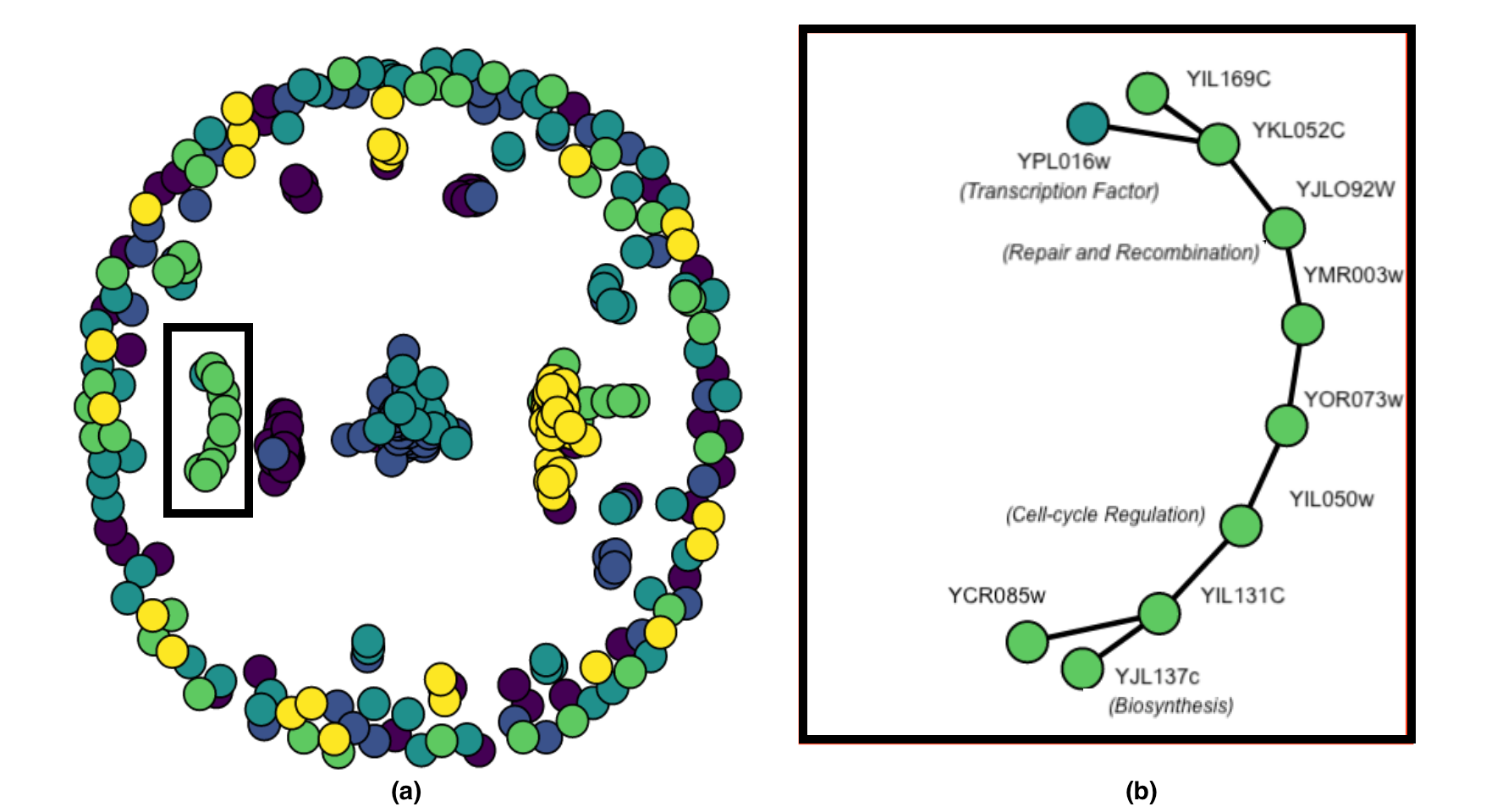}
\caption{(a) Graph representation using value-based graph construction with $delta = 90$. The position of the nodes are obtained using Fruchterman-Reingold algorithm and the color of the nodes are according to the 5 functional groupings of genes. (b) Sample of one connected component with 9 genes belonging to the 4th functional group and one gene from the 3rd functional group.}
\label{fig:vb_90}
\end{figure}

The graph representation with $\delta = 0.90$ consists the most number of non-trivial connected components. This is reflected in Figure \ref{fig:vbcc} where we have a low percentage of nodes belonging to the largest connected component and a low percentage of singletons. The graph revealed that the majority of the nodes, about $63\%$, belong to a non-trivial connected component. In Figure \ref{fig:vb_90}, we provide a visualization of the graph where we can see a total of 18 connected components and 142 singletons. Each gene is colored based on the 5 functional groups. Unlike the graph with $\delta=0.95$, here we have more connected components with genes belonging to multiple functional groups. We also show a single connected component composed mainly of genes belonging to Group 4. We highlighted some genes with known biological characterization from \cite{Cho1998} in (b) of Figure \ref{fig:vb_90}. All genes in the single component belong to the second growth phase (G2 phase) except for the transcription factor (`YPL016w') belonging to the 3rd functional group (Synthesis phase). The synthesis phase happens before the second growth phase. It is interesting however to see that genes belonging to a single connected component belong to adjacent functional groups in terms of the sequence of the cell cycle phases. 

\begin{figure}[h]
\centering
\includegraphics[width=0.48\textwidth]{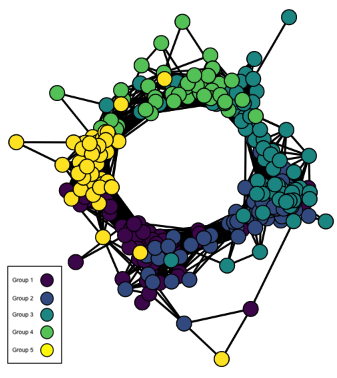}
\caption{Value-based graph construction with $\delta = 70$ }
\label{fig:vb_70}
\end{figure}

%% TODO: May suggest possible gene interaction, especially for those without further biological classification
This observation leads us to check the graph representation when we have the most number of edges. In this representation, we preserve most of the relationships but are weighted according to the correlation value. The resulting graph is shown in Figure \ref{fig:vb_70}. Analogous to the result of the NMDS visualization, we observed that there is a visual closeness of nodes belonging to a certain functional group. Moreover, the temporal characteristic of the 5 groups is also preserved. 

As a general observation, if we have a strict threshold, many weakly co-expressed genes will be disconnected. On the other hand, if we attempt to connect the weakly co-expressed genes into the network, the threshold may become so low that the genes in the strongly co-expressed network may have many links to genes belonging to a separate functional group. The next graph construction algorithm addresses this concern.

%This observation leads us to analyze the network using several community detection algorithms. The graphs generated in this section will be used in 

%Value-based graph construction contains high numbers of connected components and singletons, compared to rank-based graph construction (Figure 2) which is almost consistent at keeping its connected components and singletons at a minimum. As the singletons decrease, the connected components also decrease making a larger connected graph.

\subsection{Rank-based Graph Construction}

Similar to the previous approach, rank-based construction produces a labeled undirected graph $G(V, E)$ with the set of vertices representing each gene. Given an input parameter $d \in \mathbb{Z}^+$, rank-based construction creates edges to $d$ most similar neighbors of each vertex $u$ in $G$. The similarity metric used is also the Pearson's correlation coefficient that serves as the edge weights of the graph. We constructed a total of $10$ graphs by varying the parameter $d=\{1, \ldots, 10\}$. The details of the graphs obtained are summarized in Table \ref{table:rbnet}.

\begin{table}[h]
\resizebox{0.75\width}{!}{
\begin{tabular}{ | c | c c c c c c c c c c|}
\hline
$d$	&	1	&2	&3	&4&	5 &	6 & 7 & 8& 9& 10\\ \hline
Edges	&	323&	621 &	907&	1185	&1464	&1734 & 2020 & 2295 & 2580 & 2857\\
Singletons	&	0	&0&	0&	0&	0	&0 &0&	0&	0	&0 \\
Number of CC&	60 &	1	&1	&1&	1	&1 &1	&1&	1	&1 \\
Largest CC	& 25	&384&	384	&384&	384&	384&	384	&384&	384&	384 \\
%Clustering coefficient & 0 & 0.196 & 0.292 & 0.316 &0.353 &0.372 & 0.385 &0.4 &0.412 &0.455 \\
\hline
\end{tabular}}
\caption{Summary of network properties for graphs obtained using rank-based construction.}
\label{table:rbnet}
\end{table} 
%% TODO: Add ruynning time graph constructions

Rank-based construction guarantees that each vertex is reachable from at least $d$ vertices in the graph. As reflected in Table \ref{table:rbnet}, no singletons were created even for the graph with the least number of edges. With this construction, even with 621 edges, we can already get relationships involving all genes in the data set. In contrast with the graph obtained by value-based with $\delta=0.90$ with $1,177$ edges, we can only relate about $63\%$ of the genes in the data set. However, this construction also allows edges with weak correlation to be part of the network. 

%Studies also reported that gene co-expression networks differ from other types of biological networks in several important aspects, such as the characteristic node degree and hierarchical organization \cite{Ruan2010}, but in the work of Ruan \cite{Ruan2010}, he showed that by using rank-based graph generation, the outcome would be a graph that exhibits all the common topological properties of the other biological networks. 

\begin{figure}[h]
\centering
\includegraphics[width=0.48\textwidth]{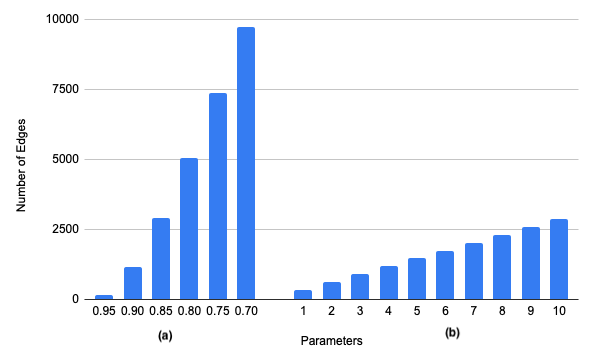}
\caption{Comparison of the total number of edges for (a) value-based construction and (b) rank-based construction.}
\label{fig:edges_vb_rb}
\end{figure}

We show the comparison of the total number of edges between all the graphs obtained by value-based and rank-based construction. In Figure \ref{fig:edges_vb_rb}, we see that the total number of edges as we increase $d$ is far below the total number of edges for the value-based construction. The highest number of edges is with parameter $d=10$ which is approximately $17\%$ of the total number of edges of the graph with $\delta = 0.70$.

\begin{figure}[h]
\centering
\includegraphics[width=0.48\textwidth]{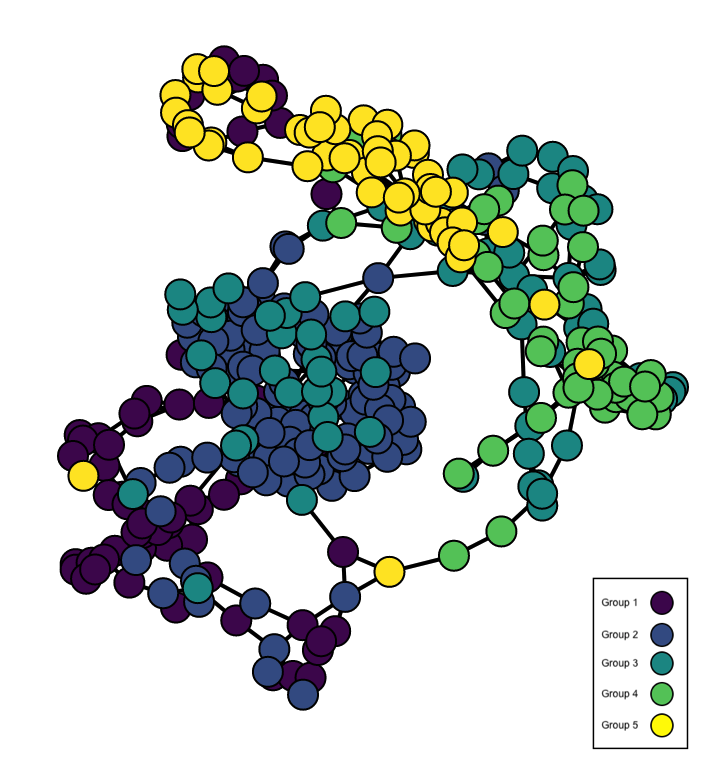}
\caption{Rank-based graph construction with $d= 2$. Colors are according to the 5 functional groups.}
\label{fig:rb_2}
\end{figure}

\begin{figure}[h]
\centering
\includegraphics[width=0.465\textwidth]{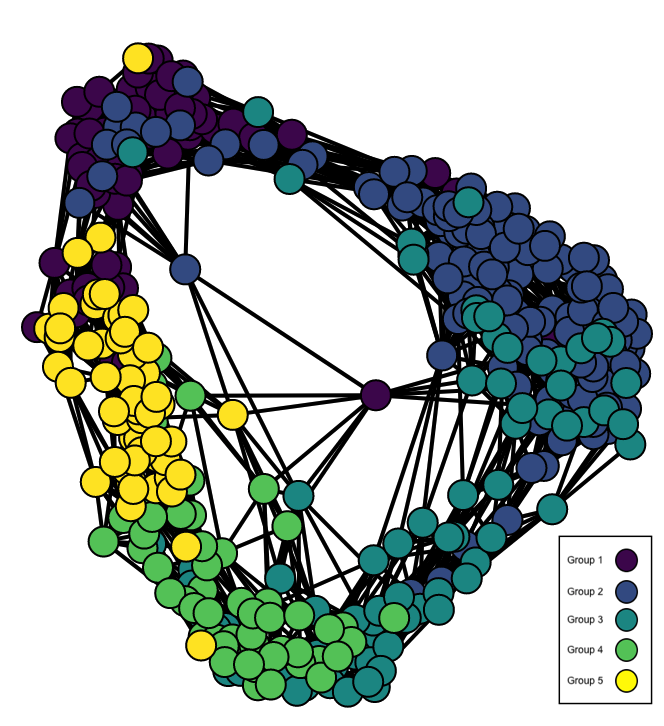}
\caption{Rank-based graph construction with $d= 10$. Colors are according to the 5 functional groups.}
\label{fig:rb_10}
\end{figure}

In Figures \ref{fig:rb_2} and \ref{fig:rb_10}, we show the two graphs obtained by rank-based construction with parameters $d = 2$ and $d=10$, respectively. We visualized both graphs using the Fruchterman-Reingold algorithm to compute for the position of the nodes and used the biological classification to color the set of nodes. With $d=2$, the resulting graph consists of all the 384 genes in the original data set. The genes belonging to groups are not visually separated as compared to the graph obtained by $d=10$. Even though the graph uses way less number of edges as compared to the graph obtained through value-based with $\delta=0.70$. The visual closeness of the nodes belonging to the same functional group is present. The temporal relationship genes that are present in the original NMDS visualization are also reflected. 

Visually, we can see the proximity of genes belonging to the same functional groups for both value-based and rank-based construction with $\delta=0.70$ and $d=10$, respectively. According to Ruan et al., \cite{Ruan2010}, the comparison of the rank-based and the value-based construction has not been rigorously examined in the literature. In the following section, we will test whether which of the following graphs can be used to predict functional groups using community detection algorithms.

\section{Community Detection}
Some graph representations in the previous section already provide an interesting group of genes belonging to an individual connected component. It is good for identifying small modules that are present in the graph but does not provide communities as partitions of the set. 

The standard approach for partitioning a data set is through clustering. The works of Ruan et al. \cite{Ruan2010} have proven that for some data sets, community detection in graphs provides more accurate partitions.  In line with this, we use $4$ different community detection algorithms to predict the grouping of genes based on the 5 functional groups. The first three algorithms use the concept of \emph{modularity} while the last algorithm uses \emph{edge-betweenness.} We'll discuss these two network metrics in the succeeding subsections. We subjected the largest component of each graph $C$ obtained by using value-based and rank-based construction with varying input parameters.  Here, we will compare how the different community detection algorithms perform in predicting the 5 functional groups. 

We use the \emph{Adjusted Rand Index} (ARI) to assess the performance of a particular community detection algorithm  in classifying the set of genes according to the biological function.  ARI is `adjusted by chance' version of the Rand Index (RI). 
\begin{equation}
ARI = (RI - \mathbb{E}(RI)) / (max(RI) - \mathbb{E}(RI))
\label{f:ari}
\end{equation}

RI is a measure of agreement between two partitioning of the same set. It is measured by getting the percentage of `agreeing' pairs with respect to the total number of pairs. A pair agree if they both belong to the same cluster or different cluster in the predicted and true clustering.

ARI has a maximum value of $1$ which means that there is a perfect agreement between the predicted and the true clustering The expected value of the ARI in the case of random clusters is 0.  A negative value of ARI suggests cluster to cluster agreement below expectation.  ARI also allows comparison between two unequal numbers of partitions.

We summarized the assessment of the predicted communities with respect to the 5 functional groupings. We displayed the ARI as well as the optimal number of clusters obtained by the greedy,  Louvain, Leiden, and Girvan-Newman algorithms in Table \ref{table:ari}. We also display some important graph properties, such as the number of nodes in the largest component $|C|$,  graph density,  and the number of edges $|E|$ for a quick reference as we compare the results.

\begin{table*}[t]

\begin{tabular}{lrrrr| rr | rr | rr | rr}
\textbf{}         & & \multicolumn{1}{l}{\textbf{}}          & \multicolumn{1}{l}{\textbf{}}        & \multicolumn{1}{l}{\textbf{}}    & \multicolumn{2}{l}{\textbf{Greedy}}       & \multicolumn{2}{c}{\textbf{Louvain}}    & \multicolumn{2}{c}{\textbf{Leiden}}     & \multicolumn{2}{c}{\textbf{Girvan-Newman}}         \\
\toprule
\multicolumn{1}{c}{\textbf{Construction}} & \multicolumn{1}{c}{\textbf{Parameter}} & $\mathbf{|C|}$& \multicolumn{1}{c}{\textbf{Density}} & \multicolumn{1}{c}{\textbf{|E|}} & \multicolumn{1}{l}{\textbf{k}} & \multicolumn{1}{l}{\textbf{ARI}} & \multicolumn{1}{l}{\textit{\textbf{k}}} & \multicolumn{1}{l}{\textbf{ARI}} & \multicolumn{1}{c}{\textit{\textbf{k}}} & \multicolumn{1}{c}{\textbf{ARI}} & \multicolumn{1}{l}{\textit{\textbf{k}}} & \multicolumn{1}{l}{\textbf{ARI}} \\ \toprule
\multirow{6}{*}{Value-Based}              & 0.70      &375     & 0.139        & 9748     & 5      & {0.232}        & 4    & 0.475    & 4    & 0.475    & 10              & 0.500    \\
       & 0.75       &369    & 0.109        & 7383     & 5      & 0.378    & 4    & \textbf{0.505}        & {4}      & \textbf{0.482}        & 10              & 0.505    \\
       & 0.80      &354     & 0.081        & 5058     & 6      & \textbf{0.400}        & 5    & 0.382    & 5    & 0.378    & 4    & 0.518    \\
       & 0.85     &310      & 0.061        & 2925     & 6      & {0.322}        & {6}      & 0.415    & 5 & {0.401}        & 5    & \textbf{0.557}        \\
       & 0.90      &116     & 0.142        & 1177     & 6      & 0.011    & 4    & 0.000    & 4    & 0.007     & 10              & 0.272    \\
       & 0.95      &53     & 0.091        & 154      & 6      & -0.025     & 7        & -0.022     & 6        & -0.023     & 4        & 0.063    \\ \midrule
\multirow{10}{*}{Rank-Based}              & 1    &25   & 0.004        & 323      & 6      & -0.011     & 5        & 0.000    & 6        & -0.011     & 10       & 0.031    \\
           & 2     &384  & 0.008        & 621      & 10     & 0.270    & 13       & 0.237    & 12       & {0.258}        & 5        & \textbf{0.499}        \\
           & 3    &384   & 0.012        & 907      & 8      & 0.329    & 10       & 0.244    & 10       & {0.278}        & 4        & \textbf{0.499}        \\
           & 4     &384  & 0.016        & 1185     & 5      & 0.356    & 8        & 0.354    & 10       & 0.290    & 5        & 0.478    \\
           & 5      &384 & 0.020        & 1464     & 5      & 0.447    & 8        & 0.302    & 8        & 0.354    & 4        & 0.493    \\
           & 6      &384 & 0.024        & 1734     & 5      & \textbf{0.479}        & {7}      & 0.332    & 8        & 0.337    & 4        & 0.493    \\
           & 7       &384& 0.027        & 2020     & 5      & 0.464    & 7        & 0.326    & 7        & {0.342}        & 4        & \textbf{0.499}        \\
           & 8      &384 & 0.031        & 2295     & 5      & 0.402    & 7        & 0.357    & 6        & 0.347    & 4        & 0.485    \\
           & 9       &384 & 0.035        & 2580     & 5      & {0.359}        & 5        & 0.360    & 5        & 0.359    & 4        & 0.492    \\
           & 10     &384 & 0.039        & 2857     & 5      & 0.372    & 5        & \textbf{0.366}        & {5}      & \textbf{0.362}        & 4        & 0.489              \\  \bottomrule
\end{tabular}
\caption{ARI of the predicted communities using different community detection algorithms for all constructed graphs.   }
\label{table:ari}
\end{table*}

\subsection{Greedy Algorithm}

Modularity is a network metric that measures the strength of the division of a network into modules or communities. It is computed by getting the number of edges within a community subtracted by the expected number of edges if the edges are placed between nodes in random order. We can compute the modularity of a given $k$-partition using 

\begin{equation}
 	Q = \sum_{c = 1}^k \left[ \frac{L_c}{|E|} - \gamma \left(\frac{d_c}{2|E|}\right)^2 \right],
 	\label{f:modularity}
\end{equation}

where $Lc$ and is the number of intra-community links in the community $c$, $d_c$ is the sum of degrees of a node in the community $c$, and $\gamma$ is the resolution parameter. %TODO: Why resolution=1 is used. We used the default resolution =1. The resolution that is less than 1 favors bigger communities while a resolution greater than 1 favors smaller communities.

The greedy algorithm is an iterative procedure that starts by treating each vertex as a separate community. In each iteration, the algorithm joins a pair of communities with the most increase in modularity until no such pair exists. The algorithm produces the best number of partitions that will maximize the modularity.

We summarize the result of predicting the communities using the greedy algorithm in Table \ref{table:ari}. For each created graph, we show the resulting ARI and the corresponding number of partitions obtained by the algorithm. The obtained the highest ARI for the graph representation was constructed using the rank-based approach with $d=5$, which is also the total number of functional groups in the data. On the other hand, we get a lower than expectation ARI with value-based and ranked-based with $\delta = 0.95 $ and $d= 1$, respectively. 

\iffalse
\begin{table} [h]
\begin{tabular}{c c c c c c }
\toprule
\textbf{Construction} & \textbf{Parameter} & \textbf{Density} & $|E|$ & $k$ & \textbf{ARI} \\ 
\midrule 
\multirow{7}{*}{Value-Based} & 0.7 & 0.139 & 9748 & 5 & 0.232 \\
 	& 0.75 	&	0.100 	&	7383	 	&5	& 0.378 \\
	& 0.8	& 0.069 	& 5058		&6 	& \textbf{0.400} \\
	& 0.85	& 0.040 	&	2925	 	& 6	&0.322\\
	& 0.9	& 0.016	 	&	1177		&6	&	0.011 \\
	& 0.95	& 0.002 &	154		&6 	&	-0.025 \\ \hline

\multirow{10}{*}{Rank-Based} &1		&0.004	 & 323	&6	&- 0.011 \\
&2		&0.008	&621	&10	&0.270 \\
&3		&0.012	&907	&8	&0.329 \\ 
&4		&0.016	&1185	&5	&0.356 \\
&5		&0.004	&464	&5	&0.447 \\
&6		&0.024	&1734	&5	&\textbf{0.479} \\
&7		&0.027	&2020	&5	&0.464 \\
&8		&0.031	&2295	&5	&0.402 \\
&9		&0.035	&2580	&5	&0.359 \\
&10		&0.039	&2857&	5	&0.372 \\
\bottomrule
\end{tabular}
\caption{ARI of the predicted communities using greedy algorithm for all constructded graphs. }
\label{tbgreedy}
\end{table}
\fi

This algorithm is also called the Clauset-Newman-Moore greedy modularity maximization with a running time of $O(|E| \cdot h \cdot \log{n})$, where $h$ is the height of the dendrogram describing the community structure. Most biological networks are sparse. As shown in Table \ref{table:ari},  we validate this claim by computing the density score of each graph created. Due to the sparsity of the graph, the parameter $h$ in the running time is bounded above by $\log{n}$. Thus, the greedy algorithm runs in $O(|E| \cdot \log^2{n})$ \cite{Clauset2004}. 
Since the total number of vertices is constant for all graphs, the running time of the algorithm for each graph is highly dependent on the total number of edges.

%Most modularity-based algorithms always produce a specific number of groups so it would be ideal to have a measure to gauge how meaningful are the divisions formed. 
%
\subsection{Louvain Algorithm} 
Louvain community detection uses a heuristic method based on modularity optimization \cite{Blondel2008}. The algorithm has two stages. The first stage is involved in creating the initial set of communities through \emph{modularity optimization}. Similar to the greedy algorithm, the first step is to assign every node to be in its community. The second step maximizes the modularity by moving each node to all of its neighbor communities. Specifically, for each node $u \in V$ and for each neighbor, $v$ of node $u$ the algorithm calculates the maximum increase in modularity for changing community $c_u$ to the community of neighbor $c_v$ across all neighbors $v$. If the modularity increases, then the algorithm move node $u$ from $c_u$ to $c_v$. Otherwise, if no positive gain is achieved the node remains in its original community. 

The second stage is called \emph{community aggregation.} In this stage, all nodes belonging to a single community are treated as a single node. Weighted connections are established by computing the sum of all the edges that traverse from one community to the other. Self-loops are present and the corresponding weight for each is obtained by getting the sum of all the edges that are present within a community. This stage is done iteratively to gain the hierarchical structure of the communities. 

Louvain algorithm also provides the optimal number of partitions based on maximum modularity. In Table \ref{table:ari},  we summarize the result of predicting the 5 functional groups using the Louvain algorithm. Likewise, we highlighted the maximum ARI for value-based and for rank-based constructed graphs. 

Louvain and the greedy algorithm produce 5 partitions for the value-based constructed graph with $\delta=0.75$. In this graph, the Louvain algorithm produces a relatively higher ARI compared to the predicted communities of the greedy algorithm. Moreover, the best ARI for the value-based exceeds any ARI from the rank-based constructed graphs. 

On the other hand, predicted communities for rank-based constructed graphs obtained a relatively lower ARI compared to the communities obtained by the greedy algorithm.

The community aggregation step of the Louvain algorithm utilizes the additional edges present in value-based construction unlike those in the rank-based constructed graphs with relatively uniform degree distribution. 
The natural topology of the value-based constructed graphs is revealed by the iterative step of the community aggregation stage of the Louvain algorithm, which is not present in the rank-based constructed graphs. 

% \usepackage{multirow}
% \usepackage{multirow}
% \usepackage{multirow}
% \usepackage{multirow}
\iffalse
\begin{table}[]
\begin{tabular}{lrrrrr}
\toprule
\multicolumn{1}{c}{\textbf{Construction}} & \multicolumn{1}{c}{\textbf{Parameter}} & \multicolumn{1}{c}{\textbf{Density}} & \multicolumn{1}{c}{\textbf{$|E|$}} & \multicolumn{1}{c}{\textbf{$k$}} & \multicolumn{1}{c}{\textbf{ARI}} \\
\toprule
\multirow{6}{*}{Value-Based} & 0.70  & 0.139 & 9748 & 5 & 0.475 \\
  & 0.75  & 0.109 & 7383 & 5 & \textbf{0.505}  \\
  & 0.80  & 0.081 & 5058 & 6 & 0.382 \\
  & 0.85  & 0.061 & 2925 & 6 & 0.415 \\
  & 0.90  & 0.142 & 1177 & 6 & 0.000 \\
  & 0.95  & 0.091 & 154 & 6 & -0.022 \\ \midrule
\multirow{10}{*}{Rank-Based} & 1 & 0.004 & 323 & 5 & 0.000 \\
  & 2 & 0.008 & 621 & 13 & 0.237 \\
  & 3 & 0.012 & 907 & 10 & 0.244 \\
  & 4 & 0.016 & 1185 & 8 & 0.354 \\
  & 5 & 0.020 & 1464 & 8 & 0.302 \\
  & 6 & 0.024 & 1734 & 7 & 0.332 \\
  & 7 & 0.027 & 2020 & 7 & 0.326 \\
  & 8 & 0.031 & 2295 & 7 & 0.357 \\
  & 9 & 0.035 & 2580 & 5 & 0.360 \\
  & 10 & 0.039 & 2857 & 5 & \textbf{0.366} \\ \bottomrule 
\end{tabular}
\label{tbLeiden}
\caption{ARI of the predicted communities using Louvain algorithm for all constructed graphs.}
\end{table} 
\fi
The running time of this algorithm is $O(n\log{n})$.  which is slightly faster compared to the running time of the greedy algorithm. 

\subsection{Leiden Algorithm}

Leiden algorithm is introduced in \cite{Traag2019} to improve the predicted communities of the Louvain algorithm.  In the Louvain algorithm, it is possible to get communities consisting of several disconnected communities.   Experimental analysis from Traag et al. \cite{Traag2019},  saw that up to 25\% of communities are `badly' connected and identified up to 16\% are disconnected.  To address this concern, the Leiden algorithm guarantees connectivity in the predicted communities.

The Leiden algorithm is a three-stage process that involves local moving of nodes,  an additional stage for refinement, and community aggregation.  The first and the third stages are synonymous with the modularity optimization and community aggregation of the Louvain algorithm.  A minor difference for the first stage,  instead of checking all nodes for membership, the Leiden algorithm only visits nodes whose neighborhood changed.  An additional stage is done after the local moving of nodes. This stage is called the refinement stage where the algorithm can further partition the communities obtained in the first stage.  The manner of splitting the communities is not done greedily but instead employs randomization.   The community aggregation step is similar to that of the Louvain algorithm.  Instead of using the initial partition of the nodes,  the Leiden algorithm uses the refined partition from the second stage. 

The Leiden algorithm guarantees that each predicted community is connected and is locally optimally assigned.  Experimental analyses showed that  Leiden outperforms the Louvain algorithm both in terms of speed and quality of the result using a subset of real-world large networks.

We also used the Leiden algorithm to predict the 5 functional groups. We listed down the optimal number of clusters as well as the corresponding ARI in Table \ref{table:ari}. We highlighted which parameter obtained the highest ARI for both value-based and rank-based constructions.

Since the Leiden algorithm is almost similar to the Louvain algorithm with small improvements in running-time and partition refinements, the comparison of the result of the community prediction is not surprising. We saw a striking similarity in the partition obtained by the Leiden and the Louvain algorithm for all value-based and rank-based graph representations.   Surprisingly,  even if the Leiden algorithm is faster than the Louvain algorithm, the ARI for both value-based and rank-based graphs is less in Leiden compared to that of Louvain.

\subsection{Girvan-Newman Algorithm}
The Girvan-Newman algorithm uses another network property called \emph{betweenness centrality} in predicting communities in a graph, unlike the first three methods that use modularity maximization. The Girvan-Newman algorithm starts from the original graph and iteratively removes edges of the graph until the final graph representation is composed of several connected components.  Each connected component is a predicted community of the algorithm. 

The manner of selecting edges to remove uses the edge property called \emph{edge betweenness}.  To simplify the definition, edge betweenness is the count of the total number of shortest paths that are passing through an edge.  The edge betweenness is used to measure the importance of a particular edge in connecting the different nodes in a graph.   Moreover,  the edges that are connecting communities are expected to have a high edge betweenness.  The idea is by removing crucial edges in the graph,  the algorithm can reveal the communities present in the original network.  

The algorithm is an iterative algorithm that starts by computing the edge betweenness of each edge that is present in the graph. The next step is to remove the edge with the highest betweenness.   Since an edge was removed,  the algorithm recomputes for the edge-betweenness of the remaining edges. The algorithm repeats this process until there are no more edges left.  The running time of the algorithm is $O(|E|^2n)$ \cite{Girvan2002}.

At the end of the computation, we are left with singletons,  where each node is a member of its own community.  The algorithm stores the connected components per iteration in a dendrogram similar to the output of hierarchical clustering algorithms.  We can extract $k$-communities depending on the level of the dendrogram. 

\begin{figure}[ht]
\centering
 \includegraphics[width=0.48\textwidth]{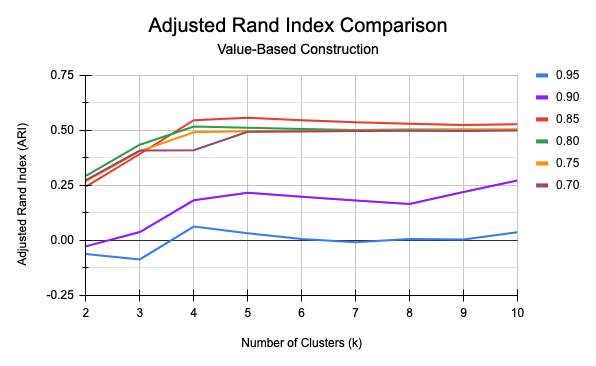}
 (a)
 \includegraphics[width=0.48\textwidth]{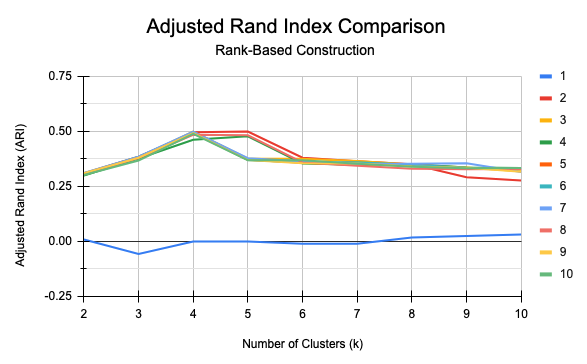}
 (b)
 \caption{Computed ARI of the value-based  and rank-based constructed graphs as we increase the total number of communities in Girvan-Newman.}
\label{fig:ari_gn}
\end{figure}

We used the Girvan-Newman algorithm for predicting communities that are present in all our graph representations.  In contrast to our first three methods,   we can specify the output to have exactly $k$ communities. With this parameter, we can compare the quality of the predicted communities as we increase the value of parameter $k$.  We summarize the results in Figure \ref{fig:ari_gn}.

We show the corresponding ARI of the predicted communities when value-based construction was used.  Each series in Figure \ref{fig:ari_gn} (a) is the ARI  as we increase the total number of communities for a graph created with a specific value of $\delta$.  Here,  we can see that both graphs with $\delta  = \{0.95, 0.90\}$ consistently obtained the lowest ARI compared to the rest of the graphs obtained using value-based construction.  Note that these two graphs have the least number of edges but with the highest correlation values.  We found an almost similar trend for the rest of the value-based constructed graphs. The ARI increases and achieves the maximum ARI at $k =5$. The trend decreases a little and plateaus as we increase $k$ up to $10$.   The graph with $\delta =0.85$ obtained the highest $ARI =0.557$  for $k=5$  when compared to the true classification of the genes.  This graph representation strikes a balance between having edges that connect strongly correlated genes and some weakly correlated genes.

In the rank-based construction, only the graph constructed with $d = 1$ obtained a 0  or even lower than expectation ARI compared to the rest of the graphs constructed using the rank-based approach.  The rest of the graphs obtained an almost similar trend.  However, as we increase $d$, the total number of edges also increases.  The running time of the Girvan-Newman algorithm is highly dependent on the number of edges as we iteratively compute for edge betweenness.  Thus making the rank-based constructed graph with $d = 2$ the best choice for predicting the 5 communities using the Girvan-Newman algorithm.  This graph representation also obtained the highest ARI = $0.499$ for $k =5$,  which is lower than that of the value-based counterpart.

In Figure \ref{fig:graph_rb_gn}, we visualize the graph obtained by rank-based construction and its corresponding predicted communities if we fix $k=5$  and as we increase the value of the parameter $d$.  Even with $d=2$,  we can already provide a prediction for all the 384 genes,  even with the least number of edges,  we can already get a good ARI of $0.499$.  The corresponding $RI=0.8174$, which means $81.74\%$ of the predicted gene pairs agree to the 5 functional groupings of genes. 

\iftrue
\begin{figure*}[ht]
\centering
 \includegraphics[width=0.8\textwidth]{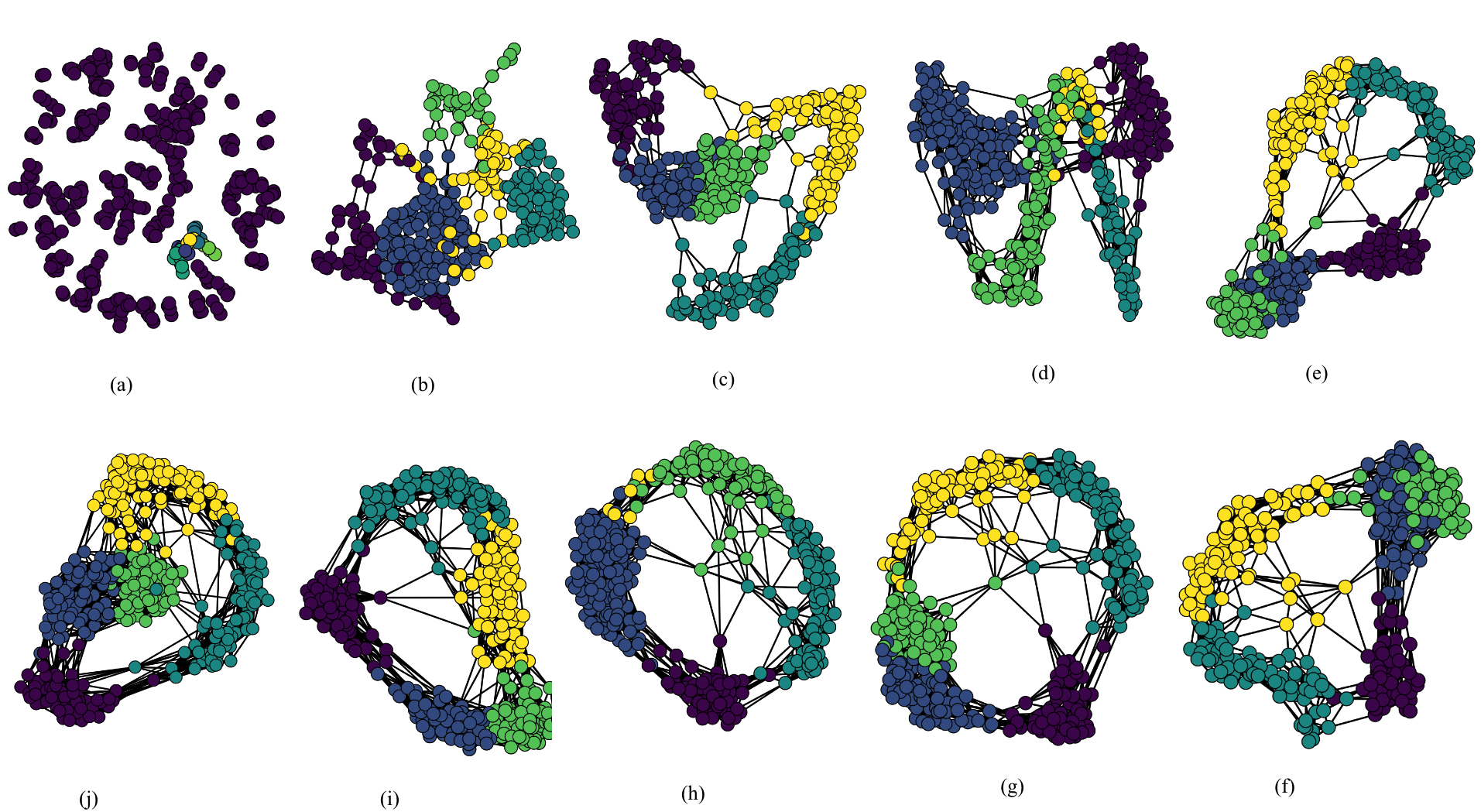}
 \caption{Predicted communities of Girvan-Newman  algorithm with $k=5$,  using all rank-based constructed graphs as we increase the parameter $d$.}
 \label{fig:graph_rb_gn}
\end{figure*}
\fi

The value-based constructed obtained a higher ARI because the total number of genes belonging to the largest component is only $310/384$ compared to the rank-based approach with the complete set of genes. The construction removes weak correlations, thus removing some genes before community detection.  The value-based construction served as a pre-processing step to remove genes that are weakly correlated with the rest of the data,  thus producing a better ARI compared to the rank-based approach.  

\begin{figure*}[ht]
\centering
 \includegraphics[width=0.8\textwidth]{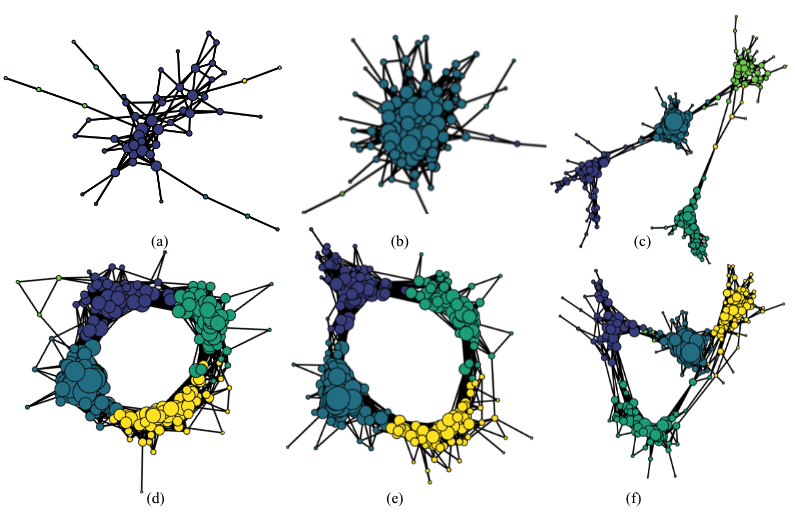}
 \caption{Predicted communities of Girvan-Newman  algorithm with $k=5$,  using all value-based constructed graph.}
 \label{fig:graph_vb_gn}
\end{figure*}

We  visualize the predicted communities of Girvan-Newman with a fix $k= 5$ for all the value-based constructed graphs.   Unlike the rank-based construction with a uniform degree distribution,  the degree of each node in a value-based constructed graph plays a crucial metric in identifying the role of each gene in a coexpression network.  In Figure \ref{fig:graph_vb_gn},  we reflected the size of each node as a function of the degree of the node.  If the number of communities is known beforehand,  we can set the parameter to $k=5$ and identified community can be used to predict the biological classification of genes with unknown functions.  

\section{Conclusion and Open Problems}

\begin{figure}[h]
\centering
 \includegraphics[width=0.48\textwidth]{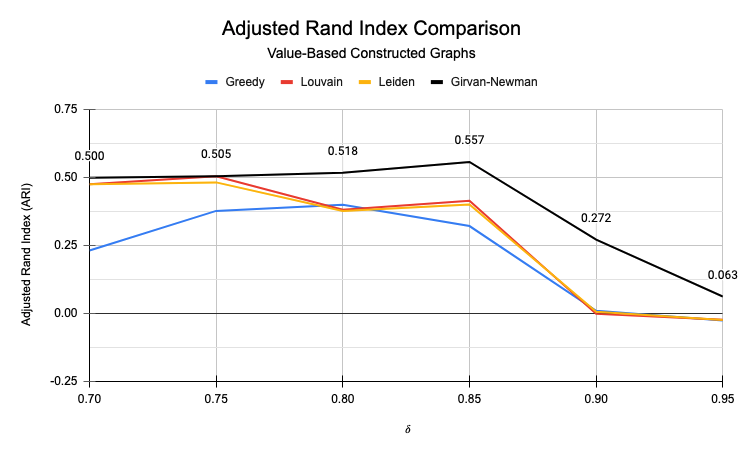}
 (a)
 \includegraphics[width=0.48\textwidth]{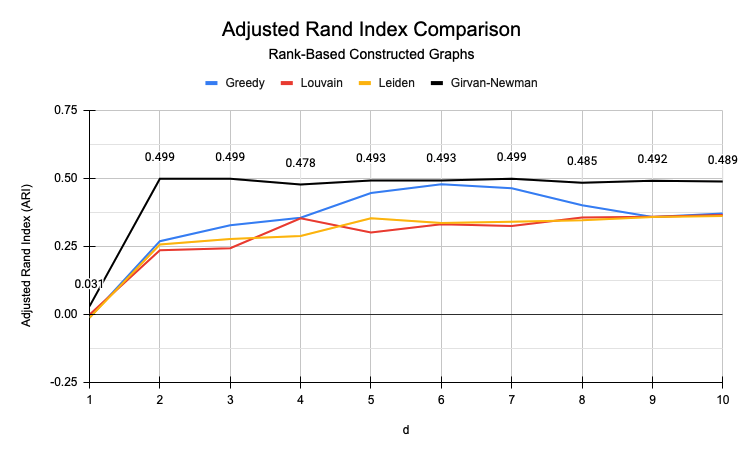}
 (b)
 \caption{Comparison of the ARI of each community detection algorithm  for all the value-based graph constructions (a) and rank-based constructions (b).}
 \label{fig:ari_summ}
\end{figure}

We compare the different community detection algorithms and identify which is better in predicting the 5 functional groups of genes.  We based our comparison using the computed ARI of each prediction.  We separate the comparison for the graphs obtained using value-based construction (a) and rank-based construction (b) in Figure \ref{fig:ari_summ}. 

Let us start with the comparison using value-based constructed graphs.  In Figure \ref{fig:ari_summ}(a), we show the performance of each algorithm as we increase the parameter $\delta$.  Note that $\delta=0.95$ consists of edges relating the strongly correlated set of genes,  while $\delta=0.70$  contains a mix of strongly and weakly co-expressed genes. The Girvan-Newman algorithm consistently obtains the highest ARI for different values of the parameter $\delta$. This was followed by the Louvain and the Leiden algorithm with an almost a similar trend.  Lastly,  as seen in the figure,  the greedy algorithm is upper bounded by the 3 other algorithms except for the graph with $\delta =0.80$.

For the rank-based constructed graphs,  the Girvan-Newman algorithm also outperforms the three modularity-based algorithms for  $d = \{1, \ldots, 10\}$.  Surprisingly, the greedy algorithm outperforms both the Leiden and the Louvain algorithm.   %TODO Answer Why
Since the $d$ is directly proportional to the number of edges of the graph and we have 3 different values of $d$  getting the maximum ARI of 0.499,  the graph constructed with $d =2$ is enough to use for the community prediction. 

We let the algorithm select the optimal number of clusters and the best number of clusters is found at $k =5$ which is the target number of communities present in the data.  Moreover, the optimal number of partitions of the modularity-based algorithm converges to 5.  

If we compare the maximum ARI obtained by the optimal number of clusters to the ARIs obtained by fixing the value of parameter $k$, the ARIs of $k=5$  is the closest for all values of $k$ that we tested for the Girvan-Newman algorithm. 

If we used the rank-based approach with $d=2$,  the total number of genes involved in the analysis is the same for the community prediction and the standard clustering algorithm.  If we fix $k=5$,  we obtained the highest  ARI with 0.499 using the Girvan-Newman algorithm, followed by $K$-means with an ARI of $0.496$, then Spectral Clustering with $0.480$,  then Birch with 0.404, and lastly Agglomerative clustering with the lowest ARI of  $0.400$.  The inclusion of the weakly correlated genes in the graph produced a closer value but still with a slightly higher ARI compared to the result of the $K$-means clustering.
 
\begin{figure}[h]
\centering
 \includegraphics[width=0.48\textwidth]{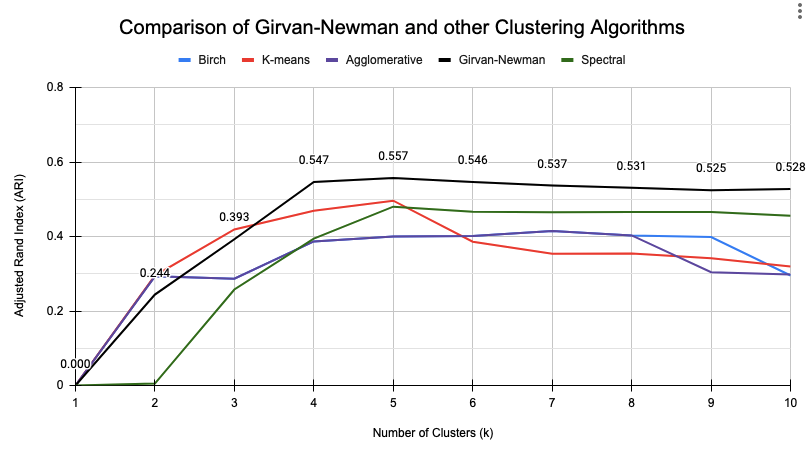}
 \caption{Comparing the ARI of the predicted communities of Girvan-Newman with the predicted clusters of  $K$-means, Spectral, Birch, and Agglomerative Clustering. }
 \label{fig:rbgnvsclustering}
\end{figure}

We are interested to verify the claim of Ruan et al. \cite{Ruan2010} that network-based methods can outperform conventional cluster analysis.   In particular,  we compare the predicted communities of the  Girvan-Newman algorithm with the result of the conventional clustering algorithms such as $K$-means,  Spectral,  Birch, and  Agglomerative clustering in Figure \ref{fig:rbgnvsclustering}.  As we increase the total number of clusters, we can see that Girvan-Newman outperforms the standard clustering algorithms in terms of the computed ARI.  The main reason for this increase is due to the filtering mechanism that is inherent in value-based constructed graphs.  Non-correlated genes are removed in the analyses making the community prediction algorithm perform better. 
In terms of running time,  the  Girvan-Newman Algorithm is the slowest among the four community detection algorithms. This is followed by the  Greedy algorithm,  Louvain, and Leiden arranged in slowest to fastest running time.  The quality of the algorithm is inversely proportional to the running time of the algorithm which is expected.

Aside from community detection, graph representation and visualization can be used for analyzing different gene interactions. We created a tool that will enable users to customize the parameters for graph creation which includes setting the graph construction algorithm to use, the definition of node, and edge weights. The user can specify the similarity metrics to use for the level of co-expression, setting node attributes, such as labels and biological classification.  We also provide an option for interactively visualizing the graph, either in a 2-dimensional or 3-dimensional view, and selecting different graph metrics or attributes for the assignment of color and node size.  This tool can be used for further studying the different relationships of genes and investigating the smaller communities present in the graph.

Some of our open problems include further investigation of the smaller connected components revealed by the value-based construction. The group of genes may suggest biological function based on their relationship with other members of the component. Since we only used the biological characterization of genes from Cho et al.\cite{Cho1998}, an immediate step involves cross-referencing the gene's classification using more updated information from biological databases such as the Kyoto Encyclopedia of Genes and Genomes (KEGG).   Another open problem of this study is to utilize the graph obtained through rank-based construction in predicting the functional group of genes with unknown functions. Moreover, we would like to investigate whether the result of our analyses will be consistent when another type of real-world network is used. 

\section{Acknowledgements}
The authors would like to thank Dr.  John Michael Yap and Dr. Henry Adorna  for their valuable inputs during presentations  in the  Algorithms and Complexity Laboratory. 

\bibliographystyle{ACM-Reference-Format}
\bibliography{pcj2022}

\end{document}